\documentclass[11pt]{revtex4}

\usepackage{graphicx}
\usepackage{epsfig}
\usepackage{latexsym}
\usepackage{amssymb}
\usepackage{bm}
\usepackage{float}
\usepackage{subfigure}
\usepackage{amsmath}
\usepackage{amsfonts}

\newcommand{\be}{\begin{equation}}
\newcommand{\ee}{\end{equation}}
\newcommand{\bea}{\begin{eqnarray}}
\newcommand{\eea}{\end{eqnarray}}

\usepackage[normalem]{ulem}

\begin{document}

\title{Gravity as the breakdown of conformal invariance}

%\date{\today}

\newcommand{\addressImperial}{Theoretical Physics, Blackett Laboratory, Imperial College, London, SW7 2BZ, United Kingdom}
\newcommand{\addressRoma}{Dipartimento di Fisica, Universit\`a La Sapienza
and Sez. Roma1 INFN, P.le A. Moro 2, 00185 Roma, Italia}

\author{Giovanni Amelino-Camelia}\email{giovanni.amelino-camelia@roma1.infn.it}\affiliation{\addressRoma}
\author{Michele Arzano} \email{michele.arzano@roma1.infn.it}\affiliation{\addressRoma}
\author{Giulia Gubitosi}
\email{g.gubitosi@imperial.ac.uk}
\affiliation{\addressImperial}
\author{Jo\~{a}o Magueijo}
\email{j.magueijo@imperial.ac.uk}
\affiliation{\addressImperial}

\begin{abstract}
We propose that at the beginning of the universe gravity existed in a limbo either because it was switched off or because it was
only conformally coupled to all particles. This picture can be reverse-engineered from the requirement that 
the cosmological perturbations be (nearly) scale-invariant without the need for inflation. It also finds support in recent results 
in quantum gravity suggesting that spacetime becomes two-dimensional at super-Planckian energies. 
We advocate a novel top-down approach to cosmology based on the idea that gravity and the Big Bang Universe are 
relics from the mechanism  responsible for breaking the fundamental conformal invariance. Such a mechanism should leave clear signatures in departures from scale-invariance in the primordial power spectrum and the level of gravity waves generated.
\end{abstract}

\maketitle

\begin{center}
{\it This essay was awarded Second Prize in the Gravity Research Foundation 2015 Awards for Essays on Gravitation}
\end{center}

\vskip 4.6cm

\newpage

Is the world simple or complicated? After decades of sustained failure in the quest for quantum gravity, one might feel sympathy for the less optimistic view. Less nervous dispositions, however, will point out that we have yet to craft a puzzle that we cannot resolve ourselves. Breakthroughs typically arise from a change of perspective, and the ability to discover simplicity in the hitherto superficially complicated problem. The unification of quantum mechanics and gravity has led to ever-more-complex and baroque constructions, but there have also been suggestions that gravity simplifies at the Planck scale. Could gravity actually switch off in the first instants of our Universe?

We are led to this conjecture by intriguing recent quantum-gravity results that hint at the fact that gravity does indeed switch off at super-Planckian energies. Most notably, the renormalization-group analysis of quantum Einstein's gravity, within the so-called ``asymptotic-safety approach", has led to speculations \cite{Krasnov:2012pd} that an ultra-violet  (UV) completion of the theory should be described by a topological theory with no local degrees of freedom. A less direct, but more general, indication favouring this picture arises from the fact that several different approaches to quantum gravity predict that spacetime becomes effectively two-dimensional in the UV limit. Evidence for this was found not only in the asymptotic-safety approach \cite{Lauscher:2005qz}, but also in Loop Quantum Gravity \cite{Modesto:2008jz}, Causal Dynamical Triangulations \cite{Ambjorn:2005db} and Ho\v{r}ava-Lifshitz (H-L) gravity \cite{Horava:2009if}. 

As we will stress below, reduction to two dimensions is known to be intimately connected to gravity switching off in the UV, so that for any equation of state perturbations evolve as in Minkowski spacetime. We may also consider the H-L model as representative of all these theories \cite{Sotiriou:2011aa,Amelino-Camelia:2013tla}, and within H-L it is known that there is a logical link between dimensional reduction to 2, gravity switching off in the ultraviolet, and the production of scale-invariant primordial density fluctuations and gravity waves  without the need for inflation. 

In this essay we argue that it could be fruitful to reverse the logic. We should {\it postulate} that at high energies the universe is gravity-free, conformally invariant and two-dimensional. Some mechanism then breaks conformal invariance and this raises the dimensionality of the universe to 4, switching on gravity. The answer to the deep questions usually investigated within quantum gravity would therefore reside in the nature of the mechanism breaking conformal invariance.  That would be where gravity ``comes from''.  It would also be where the Universe, endowed with the correct density fluctuations and gravity waves, ``comes from''.

The core feature of the H-L scenario is a modified dispersion relation (MDR) of the form\be
m^{2}=E^{2} - p^{2}(1+(\lambda p)^{2\gamma}) \, , \label{disprel}
\ee
where $E$ and $p$ are respectively the  energy and spatial momentum of a particle of mass $m$,  $\lambda$ is a small length scale (possibly the Planck length), and $\gamma$ is a parameter allowed to take integer values. It is well established \cite{Sotiriou:2011aa, Amelino-Camelia:2013tla} that for $\gamma=2$ there is running of the (spectral) dimension of spacetime from the standard value of 4 in the infrared to the reduced value of 2 in the ultraviolet.
As shown in \cite{Magueijo:2008yk,Amelino-Camelia:2013tla} for $\gamma=2$ one also finds that the vacuum quantum perturbations are scale-invariant; and they remain so as they leave the horizon and become frozen-in if Einstein gravity is assumed. This is  a general conclusion that does not depend on the background equation of state or any other details.
The assumption of Einstein gravity leads to a quadratic action for perturbations:
\be\label{action2}
S_2=\int d^{3}k\, d\eta\, a^{2}\left[\zeta'^{2}+c^{2}k^{2}\zeta^{2}\right]
\ee
where $k$ is the comoving momentum (so that the physical momentum is $p=k/a$) and $\zeta=-v/a$ is the comoving gauge curvature perturbation, with $a$ the scale factor. The resulting equation of motion for $v$ is:
\be
v''+\left[c^{2}k^{2}-\frac{a''}{a}\right]v=0 \label{EOM},
\ee
where the speed of light $c=\frac{dE}{dp}$  depends on time and on $k$, taking the UV form  $c(k,\eta)\sim \left(\frac{\lambda k}{a(\eta)}\right)^{\gamma}$.
For appropriate values of $\gamma$ the modes start inside the horizon even without inflation. For vacuum fluctuations one finds
for scales inside the horizon (up to an irrelevant phase):
\be\label{vsol}
v\sim \frac{1}{\sqrt {c k} }\sim \frac{a^{\gamma/2}}{(\lambda k)^{\gamma/2}\sqrt{k}}.
\ee
Clearly scale invariance is achieved  for $\gamma=2$, as previously announced. Importantly, this is preserved as the modes leave the horizon for all background equations of state, since $v\propto a$
outside the horizon, and this time dependence is replicated by the factor of $a$ in (\ref{vsol})  when $\gamma=2$.

At this point one might query the validity of
assuming Einstein gravity for something which is a proxy for quantum gravity. However it was shown in \cite{Amelino-Camelia:2013wha, Amelino-Camelia:2013gna} that an important insight is gained by changing units in  such a way that the dispersion relation is brought into  standard form.
This can be achieved by absorbing the time-dependence of the speed of light into a new time-unit:
\be
\tau\equiv \int c(k,\eta)d\eta\; .
\ee
Such time redefinition is $k$-dependent, a possibility which has been considered extensively in the literature, beginning with \cite{Magueijo:2002xx}, and which is often labelled as ``rainbow gravity". The $k$-dependence in the new time-unit can be removed by a redefinition of the comoving momentum trivializing the dispersion relation~\cite{Amelino-Camelia:2013gna}. This transfers the physical effects of the theory from the dispersion relation to the measure of integration in momentum space, which in the UV becomes~\cite{Amelino-Camelia:2013gna}
\be
d\mu\approx dE\, k^{\frac{2-\gamma}{1+\gamma}}\, dk.
\ee
For $\gamma=2$, this measure reflects dimensional reduction to two in the UV. Remarkably, the assumption of Einstein gravity translates in the new units into the assumption that gravity switches off. 
This is easily seen by transforming the action (\ref{action2}) and its associated equation of motion (\ref{EOM}) to the ``rainbow frame" associated with $\tau$:
\be
\frac{d^{2}v}{d\tau^{2}}+\left[k^{2}-\frac{1}{y}\frac{d^{2}y}{d\tau^{2}}\right]v=0
\ee
where  $y=a\sqrt{c}$. One can check that  for $\gamma=2$ the variable $y$ is  time-independent in the UV ($y\sim \lambda k$), so that the equation of motion for perturbations reduces to
\be
\frac{d^{2}v}{d\tau^{2}}+k^{2}v=0,
\ee
i.e. the equation one  would have in flat spacetime, or the equation valid in an expanding universe for pure radiation (which is conformally invariant and so does not feel the expansion).

No wonder our result is independent of the background equation of state. In fact gravity has dropped out of the calculation.
Within an expanding FRW universe (which is conformally equivalent to a flat Minkowski spacetime) perturbations evolve as if there were no expansion - so if perturbations have a scale-invariant spectrum inside the horizon, they will still have this property once they exit the horizon, as they are not sensitive to the presence of the horizon.

Let us now seek the general lessons that can be gleaned from our case study based on the H-L model. We found that 
in an appropriate frame, and for $\gamma=2$, three things happen concurrently: i)  UV dimensional reduction to two, ii) gravity  
conformally coupled to all particles, and iii)  pervasive scale invariance. 
In work in preparation  \cite{Vac-measure} we have part-generalized this result, proving that in a large class of  theories vastly transcending the H-L model,  dimensional reduction to 2 goes hand in hand with 
scale-invariant vacuum fluctuations.  This is true ignoring gravity.
It is therefore alluring to {\it postulate} that gravity switches off in the UV for all of these theories. This amounts to postulating that fundamentally the laws of physics are gravity-free, scale-invariant, and the world is two-dimensional. We should then seek the {\it emergence} of gravity in four dimensions as the result of whichever mechanism broke the fundamental conformal invariance.

In our case study conformal invariance is encoded in the time-independence of the variable $y$ that governs the Jeans' instability of the evolution of the perturbations. The mechanism that leads to the breaking of conformal invariance and to the emergence of gravity
 is directly connected to the requirement that in the IR the MDR governing the perturbations reduces to the trivial one. 
In the dual frame this translates into the expression for $y$:
\be
y=\lambda k\left(1+ \left(\frac{a}{\lambda k}\right)^2 \right)^{1/2}
\ee
which indeed in the IR limit reduces to $y\sim a$: gravity as we know it is thus switched on. The leading order
terms in the UV should provide insights on the mechanism breaking conformal invariance.  However, while we feel that it is likely that the qualitative aspects of our analysis apply beyond the context of our H-L object lesson, the mechanism breaking 
conformal invariance is not sufficiently general. It is tempting to observe that conformal invariance is prone to be affected by quantum anomalies~\cite{Duff:1993wm}. It would be fascinating to seek opportunities for a quantum-anomaly origin for the breakdown of conformal  invariance needed by the type of scenario we envisage, since it would lead to the remarkable hypothesis that our Universe actually originated from a quantum anomaly.

In summary cosmology appears to love the assumption that the universe starts  2-dimensional, with gravity switched off, or with all matter conformally coupled to gravity (just as happens with standard radiation subject to Einstein gravity). Then, the correct kind of fluctuations can be produced without the need to invoke inflation. In our model calculation explicit breaking of conformal invariance is inserted by hand, but 
we should seek a fundamental mechanism for breaking this invariance. Such a mechanism will be tightly constrained by the requirement that it should unleash gravity and the Big Bang universe. The only imprint from the phase where gravity was muted would be the vacuum fluctuations left by the earlier phase. The near-scale-invariance of the primordial cosmological density fluctuations could then be seen as a major observational clue informing the search for quantum gravity, and more generally our understanding of the origins of the gravitational force.  The different fundamental mechanisms for breaking conformal invariance could be tested by exploiting their relation with departures from exact scale-invariance in the power spectrum of the primordial fluctuations and  the primordial gravity waves generated.

\end{document}